\documentclass[twocolumn]{aastex701}

\usepackage[utf8]{inputenc} 
\usepackage{graphicx}
\usepackage{subfigure} 
\usepackage{booktabs}
\usepackage{multirow}

\usepackage{float}
\usepackage{threeparttable} 

\usepackage{hyperref}      
\usepackage{nameref}       
\usepackage{bookmark}      

\newcommand{\rev}[1]{#1}

\begin{document}

\title{Stellar Density Classification and Regression for CSST Multi-color Imaging Using Deep Learning}

\author[0009-0003-7031-9498]{Jinzhi Lai}
\affiliation{National Astronomical Observatories, Chinese Academy of Sciences, Beijing 100101,  People's Republic of China}
\affiliation{School of Astronomy and Space Science, University of Chinese Academy of Sciences, Beijing 100049, People's Republic of China}
\email{laijz@bao.ac.cn}

\author[]{Man I Lam$^*$}
\affiliation{National Astronomical Observatories, Chinese Academy of Sciences, Beijing 100101,  People's Republic of China}
\email{mlam@nao.cas.cn}

\author{Jianjun Chen}
\affiliation{National Astronomical Observatories, Chinese Academy of Sciences, Beijing 100101,  People's Republic of China}
\email{jjchen@nao.cas.cn}

\author[0000-0001-7314-4169]{Xin Zhang}
\affiliation{National Astronomical Observatories, Chinese Academy of Sciences, Beijing 100101,  People's Republic of China}
\email{zhangx@nao.cas.cn}

\author[0000-0003-3347-7596]{Hao Tian}
\affiliation{National Astronomical Observatories, Chinese Academy of Sciences, Beijing 100101,  People's Republic of China}
\affiliation{Institute for Frontiers in Astronomy and Astrophysics, Beijing Normal University, Beijing 100875, China}
\email{tianhao@nao.cas.cn}

\author[0009-0007-2538-5102]{Xiaohan Chen}
\affiliation{National Astronomical Observatories, Chinese Academy of Sciences, Beijing 100101,  People's Republic of China}
\affiliation{School of Physics and Astronomy, China West Normal University, Nanchong 637009, People's Republic of China}
\email{xhchen@bao.ac.cn}

\author[0000-0002-1418-9081]{Jialu Nie}
\affiliation{National Astronomical Observatories, Chinese Academy of Sciences, Beijing 100101,  People's Republic of China}
\affiliation{School of Astronomy and Space Science, University of Chinese Academy of Sciences, Beijing 100049, People's Republic of China}
\email{niejialu@nao.cas.cn}

\author[0000-0001-8247-4936]{Ming Yang}
\affiliation{National Astronomical Observatories, Chinese Academy of Sciences, Beijing 100101,  People's Republic of China}
\affiliation{Key Laboratory of Space Astronomy and Technology, National Astronomical Observatories, Chinese Academy of Sciences, Beijing 100101, People's Republic of China}
\email{myang@bao.ac.cn}

\author[0000-0002-1802-6917]{Chao Liu$^*$}
\affiliation{National Astronomical Observatories, Chinese Academy of Sciences, Beijing 100101,  People's Republic of China}
\affiliation{School of Astronomy and Space Science, University of Chinese Academy of Sciences, Beijing 100049, People's Republic of China}
\affiliation{Institute for Frontiers in Astronomy and Astrophysics, Beijing Normal University, Beijing 100875, China}
\affiliation{Zhejiang Lab, Hangzhou 311121, China}
\email{liuchao@bao.ac.cn(Lead Corresponding Author)}

\begin{abstract}
The Chinese Space Station Survey Telescope (CSST) aims to map the universe across an unprecedented dynamic range of stellar densities, spanning from extragalactic voids to the crowded Galactic center (e.g. a few stars and galaxies in the voids and $>10^5$ stars per detector in Galactic center). However, processing such heterogeneous data with a general source extraction pipeline introduces significant systematic uncertainties, standard algorithms exhibit poor accuracy in crowded fields and suffer from increased astrometric uncertainty in void regions. To mitigate these systematics, we propose a hierarchical, two-stage deep learning model for adaptive data reduction. The first stage ('classification') employs a ResNet-34 model to classify images into six discrete density categories, achieving $98.83\%$ in global accuracy. This classification acts as a critical decision gate, ensuring high calibration accuracy in the crowded fields. In the second stage ('regression'), a ResNet-50 regression model predicts the bright stars ($<23.5$ mag) in the field, which is essential for astrometric calibration, achieving a mean absolute error (MAE) of 0.0824 dex. By decoupling density characterization from source extraction, our model ensures that photometric and astrometric algorithms are optimally matched to the stellar density environment, thereby enhancing the fidelity and homogeneity of CSST as well as future large sky survey data products.
\end{abstract}


\section{INTRODUCTION}

Wide-field surveys are pivotal to modern astronomy, providing a comprehensive mapping of the universe that spans from crowded stellar fields to distant extragalactic objects. These pioneering surveys have left a huge legacy to the community through specialized campaigns. For instance, the Sloan Digital Sky Survey (SDSS; \citealt{york2000sloan}) and the Dark Energy Survey (DES; \citealt{dark2016dark, abbott2018dark}) have yielded breakthrough discoveries regarding the large-scale structure of the cosmic web. In the local universe, the \textit{Gaia} mission \citep{prusti2016gaia}, in synergy with massive spectroscopic campaigns such as LAMOST \citep{cui2012lamost, zhao2012lamost} and APOGEE \citep{majewski2017apogee}, has revolutionized our view of the complex chemodynamical history of the Milky Way. Furthermore, time-domain facilities like Pan-STARRS \citep{chambers2016panstarrs} and the Zwicky Transient Facility (ZTF; \citealt{bellm2018zwicky}) have systematically cataloged a multitude of transient phenomena. By advancing our understanding of fundamental physical processes and cosmological evolution, these surveys have established the benchmark for current astrophysical research.


Historically, large sky surveys have predefined specific sky regions during the strategy phase, based on their primary scientific goals. For example, the SDSS second generation of the Apache Point Observatory Galactic Evolution Experiment (APOGEE-2), focused on galactic archaeology and exploring the dynamical and chemical patterns of Milky Way stars, targets crowded stellar regions using the cohort strategy \citep{zasowski2017target}. In contrast, \emph{Euclid} is a significant cosmological mission, which aims at large sky coverages, targeting high galactic latitude regions to avoid crowded stellar fields \citep{laureijs2011euclid, adam2019euclid, Scaramellar2022}. Similarly, other modern missions handle crowding through specialized but static protocols. \emph{Gaia} processes crowded regions by introducing a scene-based assessment to flag blended sources, preserving catalog reliability \citep{riello2021gaia}. The \emph{Nancy Grace Roman Space Telescope} relies on specialized Difference Image Analysis (DIA) pipelines to subtract static backgrounds in the Galactic bulge \citep{penny2019predictions}. Meanwhile, the Vera C. Rubin Observatory (LSST) adopts a multi-tiered strategy: it utilizes DIA for time-domain science \citep{bosch2019overview} while employing the non-parametric deblending algorithm \texttt{Scarlet} to explicitly disentangle overlapping sources in deep static imaging \citep{melchior2018scarlet, suberlak2018lsst, ivezic2019lsst}. However, these strategies are typically pipeline-specific or hard-coded into the survey planning. Yet, the CSST continuous survey mode inevitably crosses the edges from crowded stellar regions \citep{2025SCPMA..6880402G, zhan2021wide}. The dramatic variation of stellar density in single exposure increases the systematic errors in both astrometric and flux calibrations. Specifically, as the source density approaches the confusion limit, the background noise becomes dominated by unresolved sources, which undermines the reliability of standard photometric tools like SExtractor \citep{bertin1996sextractor}. These variations may introduce flux boosting and selection biases, particularly when source blending becomes significant. Conversely, specialized iterative PSF-fitting tools (e.g., DAOPHOT; \citealt{stetson1987daophot}) are computationally prohibitive and prone to instability in sparse extragalactic areas. Given these constraints, we introduce a star-count regression before full data processing.

Recently, the rapid development of deep learning models (e.g., \citealt{krizhevsky2012imagenet, he2016deep}) has significantly improved source detection, classification, and photometry in large-scale astronomical surveys. Among these models, ResNet \citep{he2016deep} has demonstrated exceptional performance due to its ability to mitigate the vanishing gradient problem through residual connections, making it widely adopted for processing complex astronomical images. \rev{For instance, \cite{jia2020} employed a modified ResNet-50 for automatic detection and classification of astronomical targets. More recently, \cite{du2025deepap} proposed DeepAP, a two-stage deep learning framework based on ResNet-18 that adaptively predicts optimal aperture sizes for photometry, achieving high robustness in crowded stellar fields where traditional aperture methods suffer from contamination by neighboring sources. Similarly,  \cite{zhang2026} used a modified ResNet-18 to identify galaxy cluster-scale strong gravitational lenses in the DESI Legacy Imaging Surveys, effectively detecting rare arc-like features in densely crowded cluster environments.} Additionally, architectures leveraging ResNet as feature extractors, such as Mask R-CNN with ResNet-50/101 backbones, have been successfully applied to source deblending and instance segmentation in densely crowded fields, enabling precise separation of overlapping objects and improved photometric accuracy  \citep{burke2019deblending, bazzanini2025euclid}.

In order to solve the above issues from the CSST continuous survey, we develop a hierarchical two-stage model. The first stage employs a ResNet-34 classification model to classify images into six discrete density categories. This classification serves as a quality-control gate, directly route the extreme crowded fields to a specialized deblending algorithm to minimize photometric errors. The second stage utilizes a modified ResNet-50 regression model, conditioned on the identified density, to predict the count of bright stars critical for astrometric calibration. By using the robustness of classification for large-scale sorting and the precision of regression for calibration assessment, our model significantly enhances the adaptability and scientific reliability of the CSST data processing pipeline.

This paper is organized as follows. Section \ref{sec:data} describes the CSST simulation data, including the generation of synthetic catalogs and the pre-processing steps. Section \ref{sec:models} describes the architectures of the classification and regression models, along with the training strategies. Section \ref{sec:results} presents the performance evaluation of our models, discussing their accuracy and errors. Finally, Section \ref{sec:summary} is the summary.

\section{Data} \label{sec:data}
\subsection{The Chinese Space Station Survey Telescope}
The Chinese Space Station Survey Telescope (CSST) is a 2-m space telescope, designed to operate in the same orbit as the China Manned Space Station. It features a Cook-type off-axis three-mirror anastigmat optical system and is equipped with five first-generation instruments, including a survey camera capable of high-resolution wide-field imaging and slitless spectroscopy  \citep{zhan2021wide}. Over a decade-long mission, CSST is expected to survey approximately 17,500 deg$^2$ of the sky, covering a wavelength range from 2550 to 10000 \AA. The main survey camera performs imaging observations in NUV, $u$, $g$, $r$, $i$, $z$, and $y$ bands, and slitless spectroscopy in GU, GV, and GI bands. 

\begin{figure}[H]
    \centering
    \includegraphics[width=0.8\linewidth]{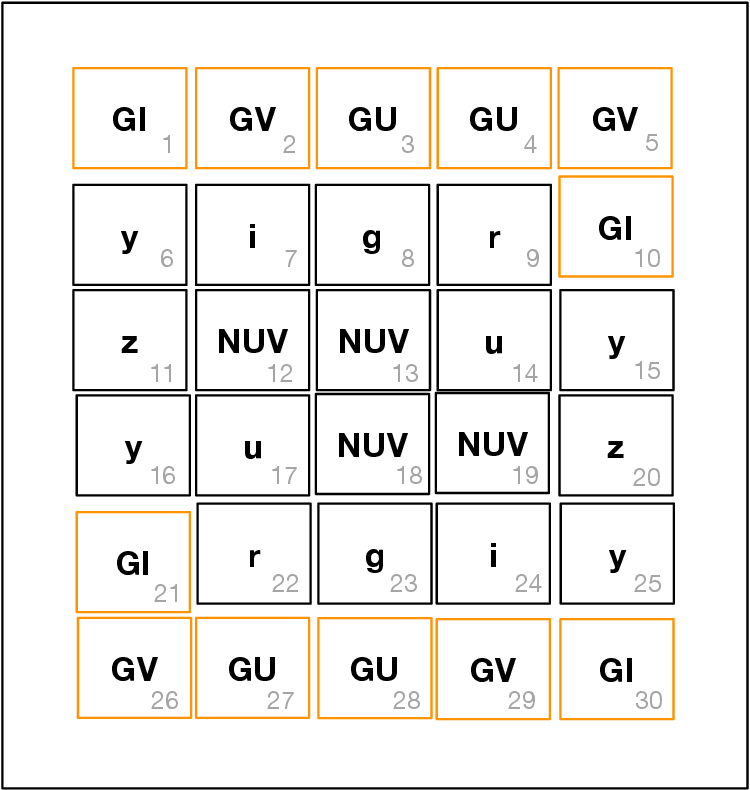}
    \caption{Schematic diagram of the main focal plane of the CSST survey camera. There are 18 black squares for detectors used in multi-color imaging observations (NUV, $u$, $g$, $r$, $i$, $z$, and $y$), and 12 orange squares for GU (255-400 nm), GV (400-620 nm) and GI (620-1000 nm) slitless spectra observations.}
    \label{fig:focal plane layout}
\end{figure}

The focal plane of the main survey camera is illustrated in Figure \ref{fig:focal plane layout}. The field of view is about $1.1^\circ \times 1.2^\circ$ in the center, which consists of 30 detectors. This study mainly focuses on the 18 detectors dedicated to multi-band imaging observations (NUV, $u$, $g$, $r$, $i$, $z$, $y$), as they provide the source data for our density estimation model, while the remaining 12 detectors are installed with slitless spectra (GU, GV, GI).

\subsection{Data Simulation}

The data used in this work were generated using the CSST simulation software, developed to produce high-fidelity, pixel-level mock data for pipeline testing and scientific evaluation  \citep{wei2025mockobservationscsstmission, ban2025mockobservationscsstmission}. The simulation mimics the telescope's image formation process, incorporating the optical system's Point Spread Function (PSF), field distortion, and key detector effects. For the classification dataset, transient artifacts (such as cosmic rays and persistence) were excluded to align with calibrated data products typically used in downstream analysis.

For the classification stage, we constructed a dataset covering the full range of stellar densities by selecting seven representative sky regions (Table \ref{tab:sky_regions}). These regions range from sparse high-latitude fields ($|b| \approx 50^\circ$) to the extremely crowded Galactic Center, which serves as a stress test for the model under high-density conditions. To represent typical Milky Way stellar fields, we selected three regions at a fixed Galactic longitude ($l=90^\circ$) but varying latitudes ($b=20^\circ, 30^\circ, 50^\circ$) to systematically sample the density gradient away from the Galactic plane.

To evaluate performance on resolved stellar populations, we explicitly incorporated the fields of the Fornax and Sculptor dwarf spheroidal galaxies (dSphs). Both are nearby, well-studied satellites of the Milky Way, providing critical benchmarks for understanding galaxy evolution through their distinct properties. Fornax is notable for its complex star formation history and significant dark matter content, whereas Sculptor is a classic, gas-poor system dominated by ancient, metal-poor stars. Their distinct properties provide complementary environments for testing our simulations and model training, similar to other high-density testbeds developed for the CSST mission \citep{xie2025mockobservationscsstmission}.

For these selected regions, we generated images in the $u, g, r, i, z, y$ bands with an exposure time of 150s. In the classification phase, the simulations were configured to include only stellar sources to isolate density features.

For the regression stage, we used a specialized dataset from the Cycle-9 (C9) simulation release to enable precise number prediction \citep{wei2025mockobservationscsstmission}. This dataset covers a 25 deg$^2$ sky region and includes both stars and galaxies to represent realistic observational conditions. It focuses specifically on the critical density range of 0 to 2000 bright stars ($< 23.5$ mag), which is essential for astrometric calibration. The C9 simulation includes comprehensive modeling of astrophysical and instrumental effects, optimized to match the expected operational environment of CSST.

\begin{deluxetable*}{lccc}
\tablewidth{0pt}
\tablecaption{Characteristics of Selected Sky Regions in the Input Catalog \label{tab:sky_regions}}
\tablehead{
\colhead{Region} & \colhead{RA Range (J2000)} & \colhead{Dec Range (J2000)} & \colhead{Stellar Density} \\
\colhead{} & \colhead{[deg]} & \colhead{[deg]} & \colhead{}
}
\startdata
Galactic\_l90\_b20 & (286.6, 291.5) & (57.8, 60.3) & $0$ -- $3{\times}10^3$ \\
Galactic\_l90\_b30 & (266.7, 271.9) & (59.8, 62.3) & $0$ -- $1{\times}10^3$ \\
Galactic\_l90\_b50 & (229.3, 233.8) & (54.8, 57.3) & $0$ -- $6{\times}10^2$ \\
Simulation Default & (236.1, 256.9) & (32.0, 48.0) & $1{\times}10^2$ -- $8{\times}10^2$ \\
Fornax & (38.4, 41.5) & (-35.8, -33.2) & $10^2$ -- $4{\times}10^3$ \\
Sculptor & (13.5, 16.6) & (-35.0, -32.5) & $10^2$ -- $10^3$ \\
Galactic Center & (264.9, 265.3) & (-28.5, -28.2) & $10^4$ -- $2{\times}10^5$ \\
\enddata
\end{deluxetable*}

\subsection{Data Preprocessing}\label{subsec:data pre}
The multi-color imaging data, generated via the CSST simulation software, comprises images with an original resolution of 9232 $\times$ 9216 pixels. To optimize computational efficiency and memory usage during data loading while maintaining compatibility with deep learning architectures, all images underwent a standardized preprocessing pipeline.

\rev{To ensure numerical stability and preserve accurate statistical properties, we implemented a mask-aware preprocessing sequence on the original high-resolution images. First, a binary mask was generated for each 9232×9216 image to identify invalid pixels, specifically those containing NaN (Not a Number) and infinity values resulting from simulated detector artifacts or saturation. To prevent mathematical failure during tensor operations, all such invalid pixels in the image channel were systematically replaced with zeros as a numerical necessity. }

\rev{Next, we applied a mask-aware per-image standardization directly to the high-resolution image. The mean ($\mu$) and standard deviation ($\sigma$) were calculated exclusively from the valid pixels identified by the mask. The image was then standardized by subtracting this mean and dividing by the standard deviation. This strategy exactly centers the background noise of valid regions around zero. Consequently, the zero-filled invalid regions act as numerically stable, neutral placeholders that align with the global mean, ensuring they do not introduce spurious signals or bias the convolutional filters.}

\begin{figure*}[htbp]
    \centering
    \includegraphics[width=\textwidth]{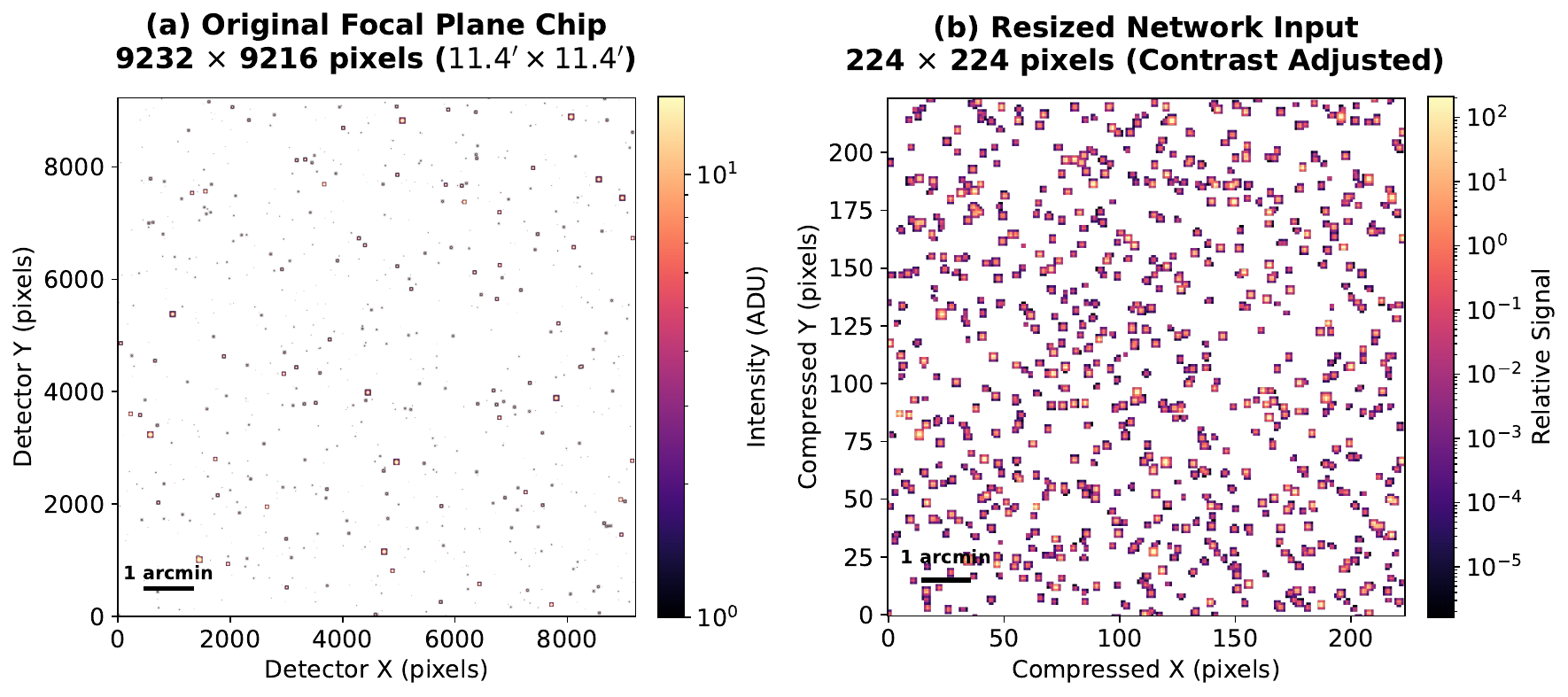}
    \caption{Visualization of the mapping between the raw CSST detector data and the network input. (a) The original high-resolution detector patch ($9232 \times 9216$ pixels, $11.4' \times 11.4'$). (b) The corresponding $224 \times 224$ feature map used as the input for the ResNet model. Both panels utilize a logarithmic stretch to enhance visibility.}
    \label{fig:resize_comparison}
\end{figure*}

\rev{The standardized image and its corresponding binary mask were then stacked to form a two-channel tensor, providing the network with explicit spatial information to distinguish physical signals from data voids. Finally, this dual-channel tensor was resized to 224$\times$224 pixels. As visualized in Figure \ref{fig:resize_comparison}, this downsampling process represents an extreme spatial compression, where the coupled mask channel ensures that the underlying density signatures remain correctly interpreted.}

\rev{It is important to clarify that the ground-truth labels for our models are derived solely from high-resolution truth catalogs based on the original 9232$\times$9216 detector coordinates. These labels remain invariant regardless of any image-level transformations. While resizing to 224$\times$224 pixels leads to significant stellar blending (crowding), the framework does not perform individual object detection. Instead, it treats density estimation as a feature-based mapping task. Even when individual point sources are blended, their integrated flux and spatial variance provide a 'statistical signature' that the ResNet architecture is trained to map back to the high-resolution physical truth. This methodology ensures that the model effectively recovers the true underlying density from compressed feature maps, maintaining scientific integrity despite the reduction in input resolution.}

For the regression task targeting bright star counts ($<$23.5 mag), \rev{we applied an additional logarithmic transformation to the labels (i.e., the count number of bright stars per image).} Since star counts are non-negative integers, we used the transformation $y' = \log_{10}(1 + y)$, where y is the original star count. This log-plus-one mapping naturally handles zero counts (giving a transformed value of zero) and compresses the skewed distribution of higher star counts, which improves training stability \citep{osborne2002notes}.

Finally, each preprocessed image and its corresponding transformed label were saved as individual .pt files using PyTorch's serialization format for efficient data handling during model training. This approach significantly reduces I/O overhead during the training process compared to reading directly from FITS files.

\rev{To enhance the diversity of the training dataset and reduce overfitting, we applied data augmentation techniques during the data loading phase \citep{shorten2019survey}. For each original image in the training set, we generated three additional augmented versions, effectively expanding the training volume by a factor of four. Each augmented copy was produced by applying a sequence of random transformations, including horizontal flipping and discrete rotations (randomly selected from 0$^\circ$, 90$^\circ$, 180$^\circ$, and 270$^\circ$). Consequently, the model was exposed to a diverse subset of the 8 possible geometric orientations across different training instances. Additionally, random color jittering (adjusting brightness and contrast) was applied to the image channel to simulate variations in background levels. This 4-fold expansion increased the classification training set to 90,240 samples and the regression training set to 101,544 samples. The validation and test sets remained un-augmented to provide a rigorous baseline for performance evaluation.}

\subsection{Dataset}
The dataset used in this study comprises two complementary subsets designed for the different stages of our density estimation framework: a classification dataset for coarse density categorization and a regression dataset for precise bright star counting.

These datasets were constructed using pixel-level simulation images and their associated truth catalogs provided by the CSST simulation pipeline. Each 9232$\times$9216 pixel image is accompanied by a catalog that records the fundamental physical properties of every simulated source, including its classification (obj\_type) and apparent magnitude (mag). The labels for our models were derived directly from these catalogs: the classification labels represent the total count of 'star' entries, while the regression labels represent the count of stars with mag $<$ 23.5.

\begin{figure*}[ht!]
    \centering
    \includegraphics[width=\linewidth]{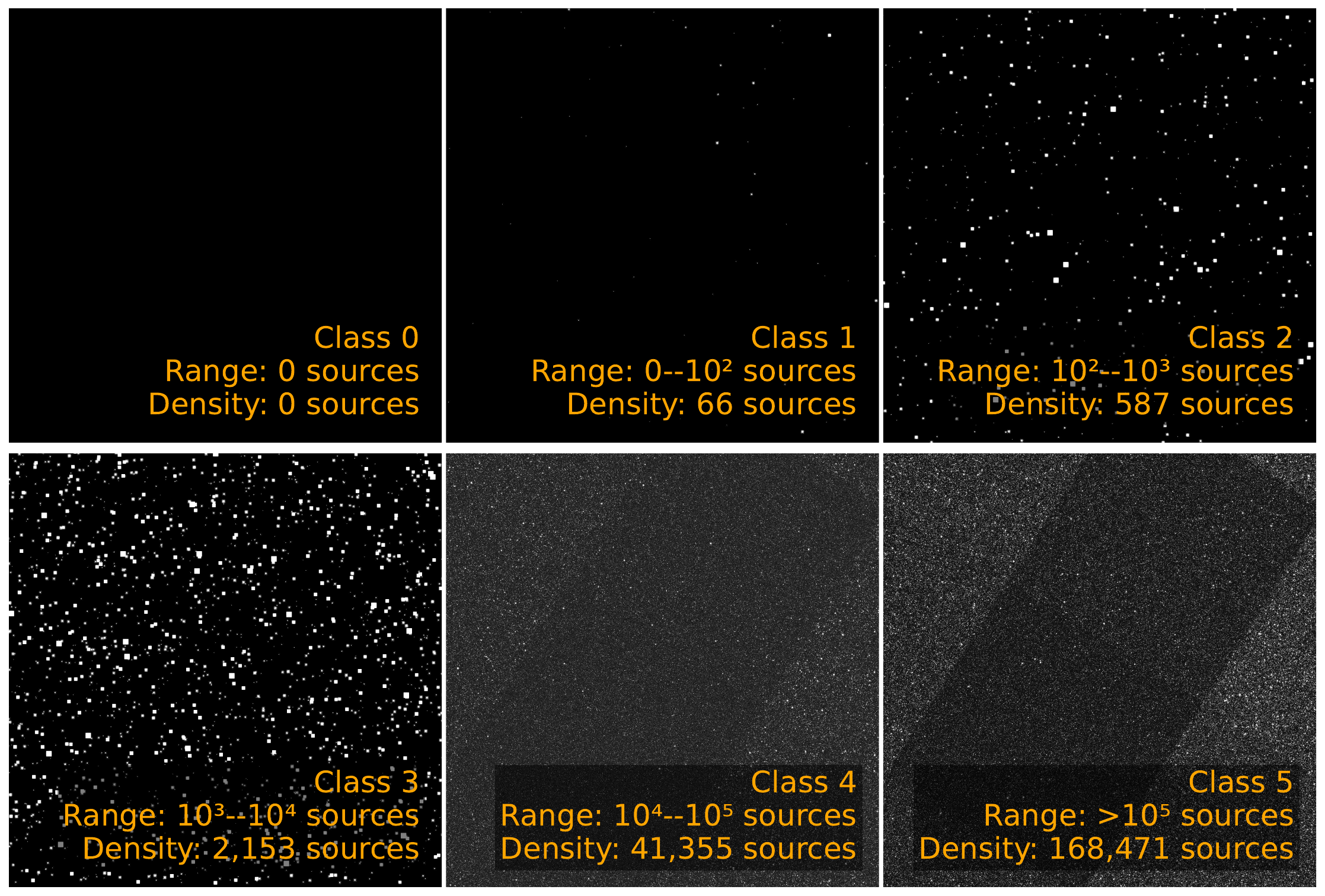}
    \caption{Representative samples of the six density classes. Each panel corresponds to the full observational area of one CSST detector chip (9232$\times$9216 pixels), demonstrating the global density patterns used for classification. The orange label indicates the density class, ranging from 0 (empty field) to 5 ($>10^5$ sources).}
    \label{fig:density_samples}
\end{figure*}

\begin{figure*}[ht!]
    \centering
    \includegraphics[width=\linewidth]{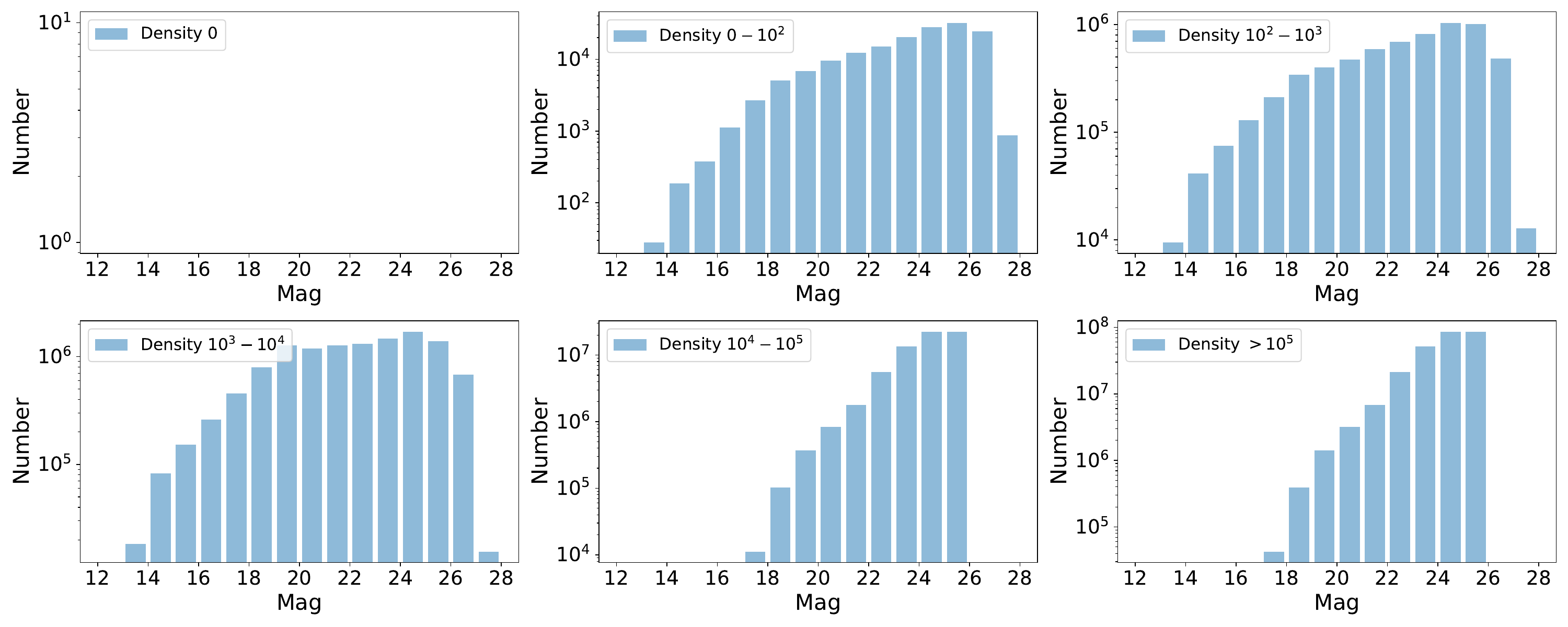}
    \caption{The magnitude distribution at different densities in the classification dataset. Note that the density range represented by the label in Title is visible in Table\ref{tab:label}. }
    \label{fig:mag_dist}
\end{figure*}

The classification dataset consists of multi-color imaging data categorized into six classes based on the number of stars in each image, with labels ranging from 0 to 5, corresponding to star counts of 0, $0-10^2$, $10^2-10^3$, $10^3-10^4$, $10^4-10^5$, and $>10^5$, respectively. This dataset contains only stellar sources, without galaxy contributions, and serves as the foundation for the initial coarse classification stage. To illustrate the characteristics of each category, we select a representative sample for each class and visualize its density distribution, as shown in Figure \ref{fig:density_samples}. The orange number in the lower right corner of each plot indicates the corresponding label. Additionally, we compute the magnitude distribution of all objects across each category to provide insight into the dataset's photometric properties, as depicted in Figure \ref{fig:mag_dist}.

\begin{figure*}[htbp]
\centering
\includegraphics[width=\linewidth]{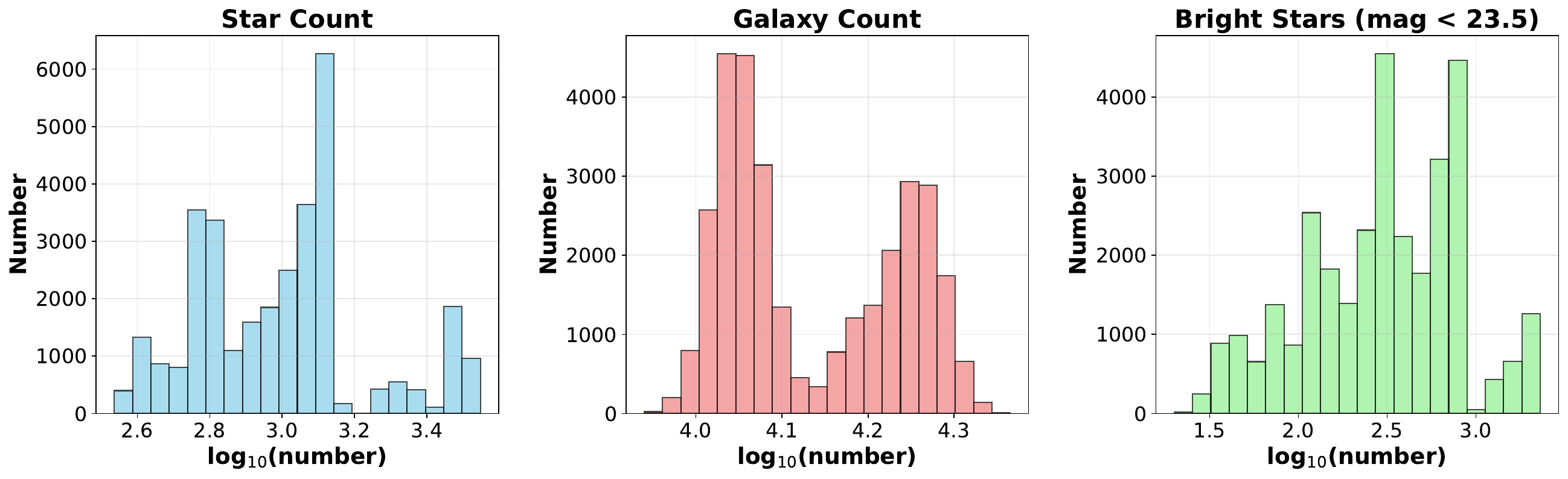}
\caption{Distribution of stellar sources, galaxy counts, and bright stars (mag $< 23.5$) in the regression dataset. The histograms show the comprehensive coverage of density variations, with particular emphasis on the bright star population essential for astrometric calibration.}
\label{fig:regression_dist}
\end{figure*}

The regression dataset, specifically designed for astrometric calibration purposes, focuses on predicting the precise number of bright reference stars ($<$23.5 mag) available for position calibration. This dataset covers a constrained but scientifically critical density range and includes both stars and galaxies to better represent real observational conditions. As shown in Figure \ref{fig:regression_dist}, the histograms present the distributions of total star counts, galaxy counts, and bright star counts across all images in this dataset. While the first two panels provide a general overview of the source composition, the third panel—showing the distribution of bright stars—directly defines the regression labels used for model training, as these bright sources are most critical for precise astrometric calibration. The bright star counts underwent a logarithmic transformation during preprocessing to address the skewed distribution and improve regression performance.

Both datasets are randomly split into training, validation, and test sets using an 8:1:1 ratio. Table \ref{tab:label} provides the detailed composition of both datasets. The classification dataset contains 22,560 training samples with 2,820 samples each in the validation and test sets. The regression dataset contains 25,386 training samples, with 3,172 validation samples and 3,178 test samples, providing substantial data for learning the continuous mapping between image features and bright star counts.

To address potential class imbalance in the classification dataset and enhance model robustness for both tasks, we apply data augmentation techniques to the training sets, as described in Section \ref{subsec:data pre}.

\begin{deluxetable*}{lcccc}
\tablewidth{0pt}
\tablecaption{Composition of Training and Testing Sets \label{tab:label}}
\tablehead{
\colhead{Dataset Type} & \colhead{Label/Range} & \colhead{Training Set (Original)} & \colhead{Validation Set} & \colhead{Test Set}
}
\startdata
Classification & 0            & 1120  & 149  & 152 \\
               & $0-10^2$     & 3324  & 433  & 411 \\
               & $10^2-10^3$  & 11050 & 1356 & 1392 \\
               & $10^3-10^4$  & 4622  & 558  & 568 \\
               & $10^4-10^5$  & 1411  & 185  & 159 \\
               & $>10^5$      & 1033  & 139  & 138 \\
\tableline
Subtotal &           & \textbf{22560} & \textbf{2820} & \textbf{2820} \\
\tableline
Regression     & 0 to over 2000 bright stars & \textbf{25386} & \textbf{3172} & \textbf{3178} \\
\enddata
\tablecomments{\rev{Sample distributions for classification and regression tasks before augmentation. To enhance model generalization, the original training sets were expanded fourfold (see Section 2.3), yielding 90,240 effective classification samples and 101,544 effective regression samples. No augmentation was applied to the validation or test sets.}}
\end{deluxetable*}

\section{Models}\label{sec:models}
Although a unified regression model could predict star counts across the full density range, we adopt a hierarchical two-stage approach to address two key limitations of such a strategy. First, photometric deblending in crowded fields requires a clear categorical trigger for switching to specialized pipelines, which classification provides more reliably than continuous regression. Second, the extreme dynamic range ($>5$ orders of magnitude) conflicts with the need for high precision in counting only bright stars ($<23.5$ mag) for astrometric calibration. A single model typically trades off accuracy in the critical low-to-moderate density regime. By separating coarse classification (ResNet-34) from targeted bright-star regression (ResNet-50), we achieve both robust pipeline routing and precise astrometric assessment.

\subsection{Model Architecture}
\subsubsection{Residual Network Foundation}
ResNet (Residual Network) is a seminal deep learning architecture introduced by  \citep{he2016deep} that enables the training of very deep neural networks while mitigating vanishing gradient problems. The core innovation is the residual block, which learns residual mappings F(x) = H(x) - x instead of direct mappings H(x), where x represents the input to the block. This is achieved through skip connections that bypass one or more layers, allowing the network to learn the difference between input and desired output rather than the complete transformation.

\begin{figure}[ht!]
    \centering
    \includegraphics[width=0.55\linewidth]{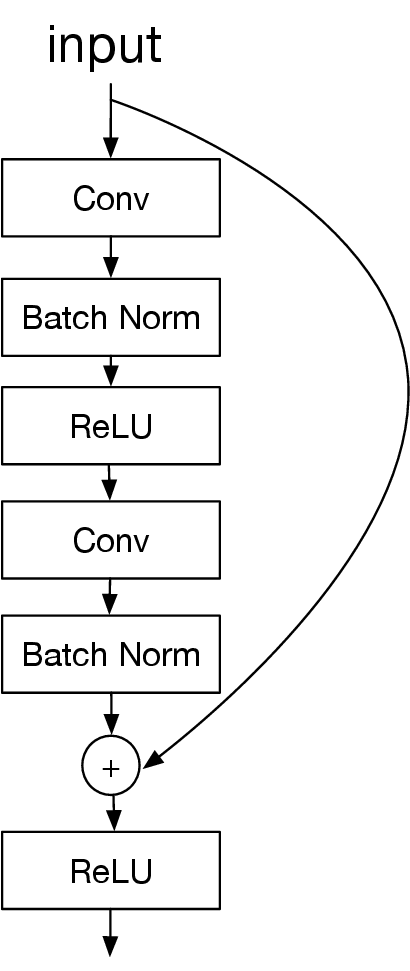}
    \caption{The structure of the ResNet block.}
    \label{fig:block}
\end{figure}

As shown in Figure \ref{fig:block}, each residual block contains multiple convolutional layers with a shortcut connection that adds the block's input to its output. This structure facilitates gradient flow during backpropagation and enables training of substantially deeper networks. Batch normalization is applied after each convolutional layer to reduce internal covariate shift, accelerate convergence, and enhance generalization \citep{ioffe2015batch}. Instead of fully connected layers, ResNet employs global average pooling to reduce feature maps to single values before the final output layer, decreasing overfitting and computational complexity.

\subsubsection{Classification Model: ResNet-34}
For coarse density classification, we employ a ResNet-34 architecture that balances computational efficiency and representational capacity \citep{he2016deep}. The model is designed to process dual-channel inputs (224$\times$224$\times$2), consisting of the simulated imaging data and its corresponding mask channel. The model begins with an initial 7$\times$7 convolutional layer producing 64 output channels, followed by max pooling and four residual layers with progressively increasing output channels (64, 128, 256, and 512). Each residual layer contains multiple basic blocks with two 3×3 convolutional layers, batch normalization, and ReLU activations.

The network concludes with global average pooling and a fully connected layer producing probability distributions across six density classes (0 to over $10^5$ stars per image). \rev{Table \ref{tab:arch}} summarizes the complete architecture, showing how the spatial resolution decreases while channel depth increases through the network, ultimately producing class probabilities through a final fully connected layer with softmax activation.

\subsubsection{Regression Model: ResNet-50}
For precise bright star counting, we utilize a deeper ResNet-50 architecture pre-trained on ImageNet, modified for our astronomical regression task \citep{he2016deep}. Key adaptations include: (1) modifying the initial convolutional layer to accept dual-channel input—comprising the science image and the data quality mask—by adjusting the input kernel dimensions while preserving pre-trained features where applicable; (2) substituting the classification head with a custom regression head featuring multiple fully connected layers (512, 256, 128, and 1 neurons) with batch normalization, ReLU activation, and dropout for regularization \citep{ioffe2015batch, nair2010rectified,srivastava2014dropout}.

The regression-specific architecture maintains the powerful feature extraction capabilities of the pre-trained ResNet-50 backbone while specializing the output layers for continuous value prediction. The final layer outputs a single value representing the predicted number of bright stars ($<$23.5 mag), enabling precise astrometric calibration.

This complementary architecture strategy provides computational efficiency: the lighter ResNet-34 enables rapid coarse classification for pipeline routing, while the more powerful ResNet-50 delivers precise regression estimates within identified density regimes.

\begin{deluxetable*}{lccc}
\tablewidth{0pt}
\tablecaption{ResNet Model Architectures \label{tab:arch}}
\tablehead{
\colhead{Layer} & \colhead{ResNet-34 (Classification)} & \colhead{ResNet-50 (Regression)} & \colhead{Output Size}
}
\startdata
Input & \multicolumn{2}{c}{Image + Mask} & 224$\times$224$\times$2 \\
Initial Conv & 7$\times$7, 64, stride 2 & 7$\times$7, 64, stride 2 & 112$\times$112$\times$64 \\
Max Pool & 3$\times$3, stride 2 & 3$\times$3, stride 2 & 56$\times$56$\times$64 \\
Residual Layer 1 & 3$\times$3, 64 $\times$3 & 1$\times$1, 64 $\times$3 & 56$\times$56$\times$64 \\
                 &                          & 3$\times$3, 64            &               \\
                 &                          & 1$\times$1, 256           &               \\
Residual Layer 2 & 3$\times$3, 128 $\times$4 & 1$\times$1, 128 $\times$4 & 28$\times$28$\times$128 \\
                 &                           & 3$\times$3, 128           &                \\
                 &                           & 1$\times$1, 512           &                \\
Residual Layer 3 & 3$\times$3, 256 $\times$6 & 1$\times$1, 256 $\times$6 & 14$\times$14$\times$256 \\
                 &                           & 3$\times$3, 256           &                \\
                 &                           & 1$\times$1, 1024          &                \\
Residual Layer 4 & 3$\times$3, 512 $\times$3 & 1$\times$1, 512 $\times$3 & 7$\times$7$\times$512 \\
                 &                           & 3$\times$3, 512           &               \\
                 &                           & 1$\times$1, 2048          &               \\
Global Pool & \multicolumn{2}{c}{Average Pool} & 1$\times$1$\times$512/2048 \\
Output & FC, 6 classes & FC, 1 value & 6/1 \\
\enddata
\tablecomments{The table compares the architectures of ResNet-34 and ResNet-50 used for classification and regression tasks, respectively.}
\end{deluxetable*}

\subsection{Training process}
\subsubsection{Classification Model Training}
We trained the classification model using Cross-Entropy loss and the AdamW optimizer. This setup effectively penalizes misclassifications by measuring the discrepancy between predicted probabilities and ground-truth labels, thereby refining the decision boundaries across the six density classes.

The AdamW optimizer was selected for its adaptive learning rate capabilities and improved weight decay handling compared to traditional SGD \citep{loshchilov2017decoupled}. To enhance training stability, we implemented gradient clipping with a maximum L2 norm of 1.0, preventing exploding gradients and ensuring stable weight updates. Additionally, an early stopping mechanism was employed to terminate training if the validation loss failed to improve after 15 consecutive epochs, effectively reducing overfitting risk and optimizing computational resources.

\subsubsection{Regression Model Training}
The regression model employed a specialized training strategy using a staged approach with layer-wise learning rate differentiation. We designed a custom combined loss function that integrates Mean Squared Error (MSE), Mean Absolute Error (MAE), and Relative MSE terms to handle the wide dynamic range of bright star counts while maintaining training stability. The loss function is defined as:

$$L = \alpha \cdot \mathrm{MSE} + \beta \cdot \mathrm{MAE} + \gamma \cdot \mathrm{RelativeMSE}$$

where \(\alpha,\beta,\gamma\) are weighting coefficients that balance the contribution of each loss component, and the Relative MSE term includes value clamping (max\_value=10.0) to prevent excessive loss values from dominating the optimization \citep{huber1992robust}.

The training process consisted of three distinct phases with progressively unfrozen layers:

Stage 1: Only the fully connected layers were trained for 30 epochs

Stage 2: Layer 4 and fully connected layers were trained for 40 epochs

Stage 3: All layers were fine-tuned for 70 epochs

The optimal base learning rate and layer-specific learning rate multipliers were determined through systematic hyperparameter optimization using the Optuna framework, ensuring optimal convergence across all training stages \citep{akiba2019optuna}. The AdamW optimizer was used throughout all stages, with ReduceLROnPlateau schedulers automatically adjusting learning rates based on validation loss plateaus.

This progressive unfreezing strategy enabled stable convergence while leveraging the pre-trained ResNet-50 feature extraction capabilities, with the optimal learning rate configuration specifically tailored to our astronomical regression task through rigorous experimental evaluation.

\subsubsection{Shared Training Infrastructure}
Both models utilized mixed-precision training (AMP) with torch.float32 precision to accelerate computation while maintaining numerical stability \citep{micikevicius2017mixed}. Hyperparameters including initial learning rates, batch sizes, and training durations were systematically optimized through extensive experimentation. The training process was monitored through loss and accuracy curves, with the best model weights preserved based on validation performance.

\begin{figure}[!ht]
    \centering
    \includegraphics[width=\linewidth]{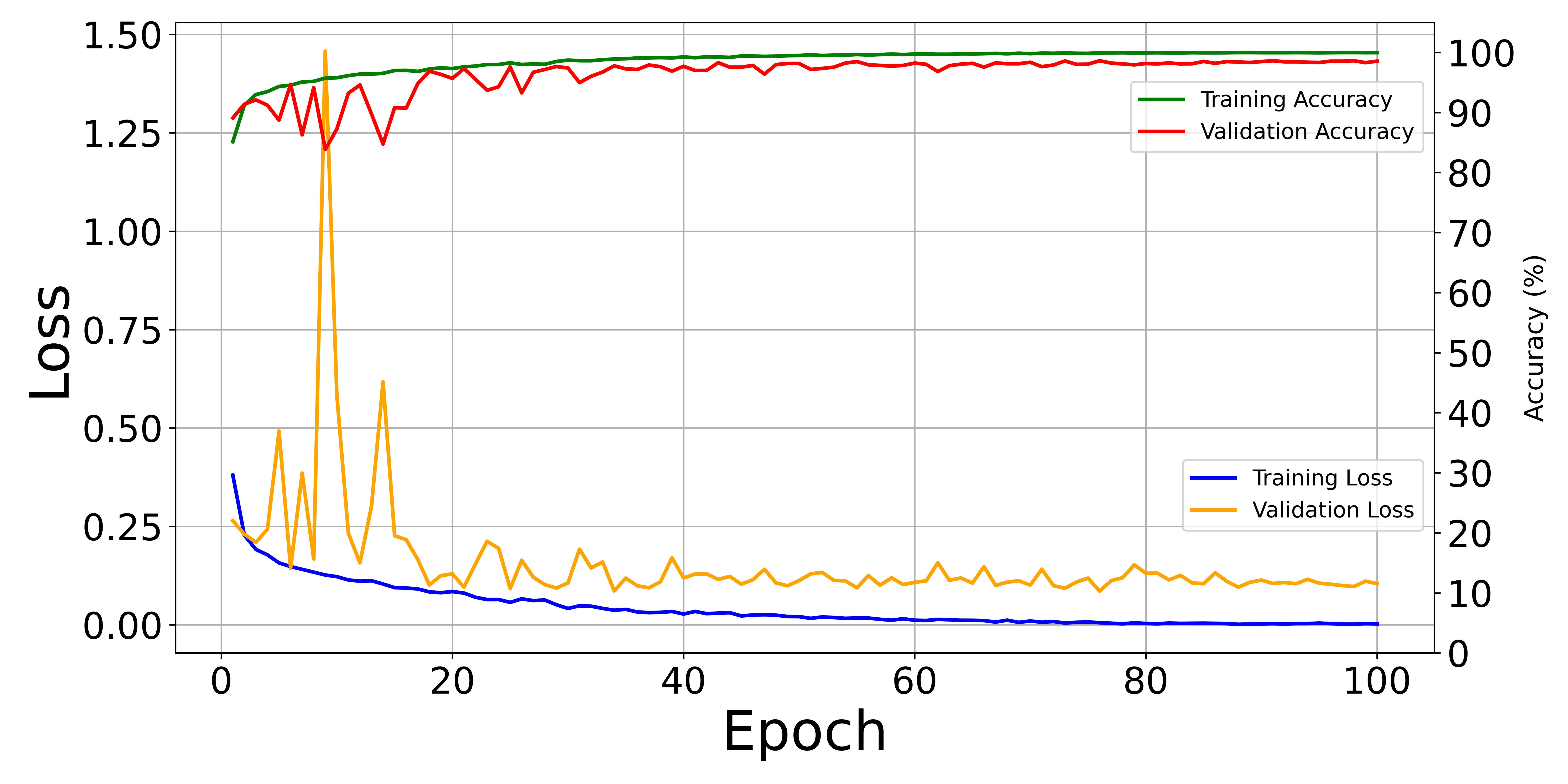}
    \caption{The final loss curve illustrating the training and validation losses over epochs. The smooth convergence indicates that the model effectively learned the underlying patterns in the data, with the training loss decreasing consistently and the validation loss stabilizing, suggesting no signs of overfitting.}
    \label{fig:loss_curve}
\end{figure}

Training convergence was characterized by smoothly decreasing training loss with stabilized validation loss, indicating effective pattern learning without overfitting. Figure \ref{fig:loss_curve} illustrates representative training curves showing this convergence behavior for both classification and regression tasks.

\subsection{Model Selection}
We evaluated several representative convolutional neural network (CNN) architectures to identify the optimal backbone for the CSST density classification task, including ResNet-34 \citep{he2016deep}, VGGNet-16 \citep{simonyan2014very}, DenseNet-121 \citep{huang2017densely}, AlexNet \citep{krizhevsky2012imagenet}, and a custom baseline CNN. The selection process was guided by a holistic engineering perspective, balancing scientific precision (Top-1 Accuracy and Macro F1) against operational efficiency (Throughput and VRAM consumption). Table \ref{tab:model_comparison} summarizes these multi-dimensional performance metrics.

\rev{As demonstrated in the comparative analysis, ResNet-34 emerged as the optimal 'mission-ready' model, offering the most robust balance between accuracy and resource utilization. While DenseNet-121 is characterized by excellent parameter efficiency (only 4.97~M parameters), our tests revealed a significant computational bottleneck: its dense connectivity and feature concatenation operations lead to the highest VRAM usage (4063~MB) and a relatively high latency of 6.56~ms/image. In the context of CSST's high-throughput requirements, such memory and time overhead would limit the parallelization capacity of the data reduction pipeline.}

\rev{In contrast, ResNet-34 utilizes its residual learning paradigm—characterized by skip connections that mitigate vanishing gradients—to achieve near-peak accuracy (98.83\%) with a highly efficient throughput of 2.77~ms/image. This throughput is approximately three times faster than that of AlexNet (7.92~ms/image), despite ResNet-34 having a deeper architecture. This superiority is primarily due to ResNet-34's more efficient parameterization; it achieves a 4.54\% higher accuracy than AlexNet while utilizing only 22\% of AlexNet's parameter count (13.67~M vs. 61.09~M).}

\rev{Furthermore, although the lightweight Baseline CNN and AlexNet require less VRAM, their significantly lower accuracies (90.04\% and 94.29\%, respectively) pose a risk to scientific integrity. The practical significance of this accuracy gap is underscored by the operational scale of the CSST mission. Based on the mission profile, CSST is designed to perform approximately 300 exposures per day, with 18 detector-sized imaging products per exposure, totaling over $1.9 \times 10^6$ images per year \citep{gong2026introduction}. At this scale, the 4.54\% accuracy advantage of ResNet-34 over AlexNet prevents nearly 90,000 instances of misclassification annually. Since classification serves as the primary 'routing gate' for subsequent specialized processing branches (e.g., deblending or PSF modeling), such errors would propagate as systematic biases in the final stellar density catalogs. By reducing the error rate from 5.71\% (AlexNet) to 1.17\% (ResNet-34), we effectively suppress these systematic risks while maintaining an operational throughput (2.77~ms/image) that easily accommodates the mission's high-volume data stream.}

\rev{VGGNet-16, despite achieving a competitive accuracy of 97.41\%, was excluded due to its excessive redundancy. With 135.4 million parameters and a high VRAM footprint (3350~MB), it imposes unnecessary computational costs without exceeding the performance of the more compact ResNet-34. Therefore, by adopting ResNet-34, we ensure the highest scientific reliability for the CSST mission while maintaining the operational efficiency required for petabyte-scale data processing.}

For the regression task, we employed ResNet-50 based on its established performance in similar regression applications and the need for a powerful feature extractor to handle complex patterns in astronomical images \citep{he2016deep}. Recent studies have demonstrated ResNet-50's effectiveness in astronomical regression tasks, including  photometric redshift estimation \citep{hayat2021self, sandeep2021analyzing} and metallicity prediction \citep{monti2024leveraging}. The deeper architecture and larger capacity of ResNet-50 provide the necessary representational power for precise bright star counting, while its pre-trained weights on ImageNet offer robust feature extraction capabilities that transfer effectively to our astronomical domain.

The key to the success of the ResNet architecture is its introduction of 'shortcut connections', or residual learning modules. This innovative design allows gradients to flow directly through deeper network layers during backpropagation, effectively mitigating the vanishing gradient problem common in deep networks \citep{ebrahimi2021study}. Therefore, ResNet can build deeper networks without sacrificing training stability, while avoiding the massive parameter burden of VGGNet and the dense connection computational overhead of DenseNet. The depth of ResNet-34 is sufficient to achieve high classification accuracy, while its computational requirements remain within a manageable range, making it an ideal compromise. The original ResNet paper explicitly notes that simply increasing network depth does not invariably lead to better performance, as excessively deep networks can encounter optimization difficulties and degradation problems \citep{he2016deep}. Furthermore, subsequent studies have demonstrated that moderately deep ResNets (e.g., ResNet-34) can outperform deeper networks on specific classification tasks with significantly lower computational costs \citep{wu2019wider, gao2021transfer, xu2024advances}. Therefore, the choice of ResNet-34 is not merely based on its highest score on the test set but is a strategic decision driven by the practical operational constraints of large astronomical survey projects. For regression, ResNet-50 provides the depth and capacity required for precise continuous value prediction. This systematic evaluation confirms that the ResNet architecture family offers an optimal balance of accuracy, efficiency, and generalizability for astronomical datasets characterized by high-resolution imaging and inherent class imbalance.

\begin{deluxetable*}{lccccc}
\tablewidth{0pt}
\tablecaption{Performance Comparison of Different Architectures \label{tab:model_comparison}}
\tablehead{
\colhead{Model} & \colhead{Top-1 Acc.} & \colhead{Macro F1} & \colhead{Params} & \colhead{Throughput} & \colhead{VRAM} \\
\colhead{} & \colhead{(\%)} & \colhead{(\%)} & \colhead{(M)} & \colhead{(ms/img)} & \colhead{(MB)}
}
\startdata
ResNet-34 (Ours) & 98.83\% & 96.27\% & 13.67  & 2.77 & 1124 \\
VGGNet-16        & 97.41\% & 97.41\% & 135.40 & 3.54 & 3350 \\
DenseNet-121     & 95.99\% & 95.99\% & 4.97   & 6.56 & 4063 \\
AlexNet          & 94.29\% & 94.28\% & 61.09  & 7.92 & 597  \\
Baseline CNN     & 90.04\% & 90.04\% & 0.42   & 0.36 & 1039 \\
\enddata
\tablecomments{Comparison of classification performance and computational efficiency across various neural network architectures.  Throughput represents the total average processing time per image during testing (including I/O and batch processing).}
\end{deluxetable*}


These results confirm that ResNet's residual learning paradigm is particularly suited for astronomical density classification tasks, where preserving faint density signatures and handling severe class imbalance are paramount. The comparative analysis suggests that while newer architectures like DenseNet achieve comparable accuracy, ResNet remains preferable for CSST applications due to its better memory efficiency and interpretable feature hierarchy.

\section{Result \& Discussion}
\label{sec:results}

\subsection{Classification Sensitivity}
The classification demonstrates robust performance across all six density classes, achieving an overall accuracy of 98.83\% with balanced precision-recall metrics as detailed in Table \ref{tab:complete_performance}. Significantly, the model demonstrates exceptional performance at density extremes: Class 0 (void fields) attains perfect precision, recall, and F1 score (100\%), while Class 5 ($> 100$k sources) maintains exceptional detection capability (99.62\% recall) alongside high prediction confidence (99.62\% precision). This performance is attributed to distinct morphological signatures—pristine background separation in sparse fields versus characteristic source crowding patterns in high-density regions.

\begin{deluxetable*}{llcccc}
\tablewidth{0pt}
\tablecaption{Complete Model Performance Metrics \label{tab:complete_performance}}
\tablehead{
\colhead{Model} & \colhead{Class/Range} & \colhead{Accuracy/MAE} & \colhead{Precision/MSE} & \colhead{Recall/RMSE} & \colhead{F1 Score}
}
\startdata
Classification & Class 0 & 100.00\% & 100.00\% & 100.00\% & 100.00\% \\
               & Class 1 & 98.74\%  & 99.37\%  & 98.74\%  & 99.05\%  \\
               & Class 2 & 98.79\%  & 99.09\%  & 98.79\%  & 98.94\%  \\
               & Class 3 & 97.90\%  & 96.33\%  & 97.90\%  & 97.11\%  \\
               & Class 4 & 99.19\%  & 99.46\%  & 99.19\%  & 99.32\%  \\
               & Class 5 & 99.62\%  & 99.62\%  & 99.62\%  & 99.62\%  \\
\tableline
Regression     & global  & 0.0824   & 0.0178   & 0.1333   & ---      \\
\enddata
\tablecomments{Performance metrics for both classification and regression tasks. For classification, metrics are presented per class; for regression, global error metrics (MAE, MSE, and RMSE) are reported.}
\end{deluxetable*}

From a scientific perspective, the high recall rate for Class 5 (99.62\%) is particularly significant. Class 5 represents the most crowded fields (e.g., Galactic Center) where standard aperture photometry fails severely due to source blending. The model's ability to reliably identify these fields ensures that they are correctly routed to specialized deblending pipelines (e.g., PSF fitting), thereby preventing crowding-induced systematic errors from contaminating the final catalog.

Class 3 (1k-10k sources) exhibits relative performance degradation (F1=97.11\%), primarily attributable to boundary concentration effects where samples cluster near the 1k-source classification threshold. We systematically analyze this boundary phenomenon in Section 4.2 through density distribution mapping and error pattern decomposition. The model misclassified 33 samples, with errors predominantly occurring at class boundaries due to subtle morphological overlaps.

\begin{figure*}[htbp]
\centering
\includegraphics[width=0.6\linewidth]{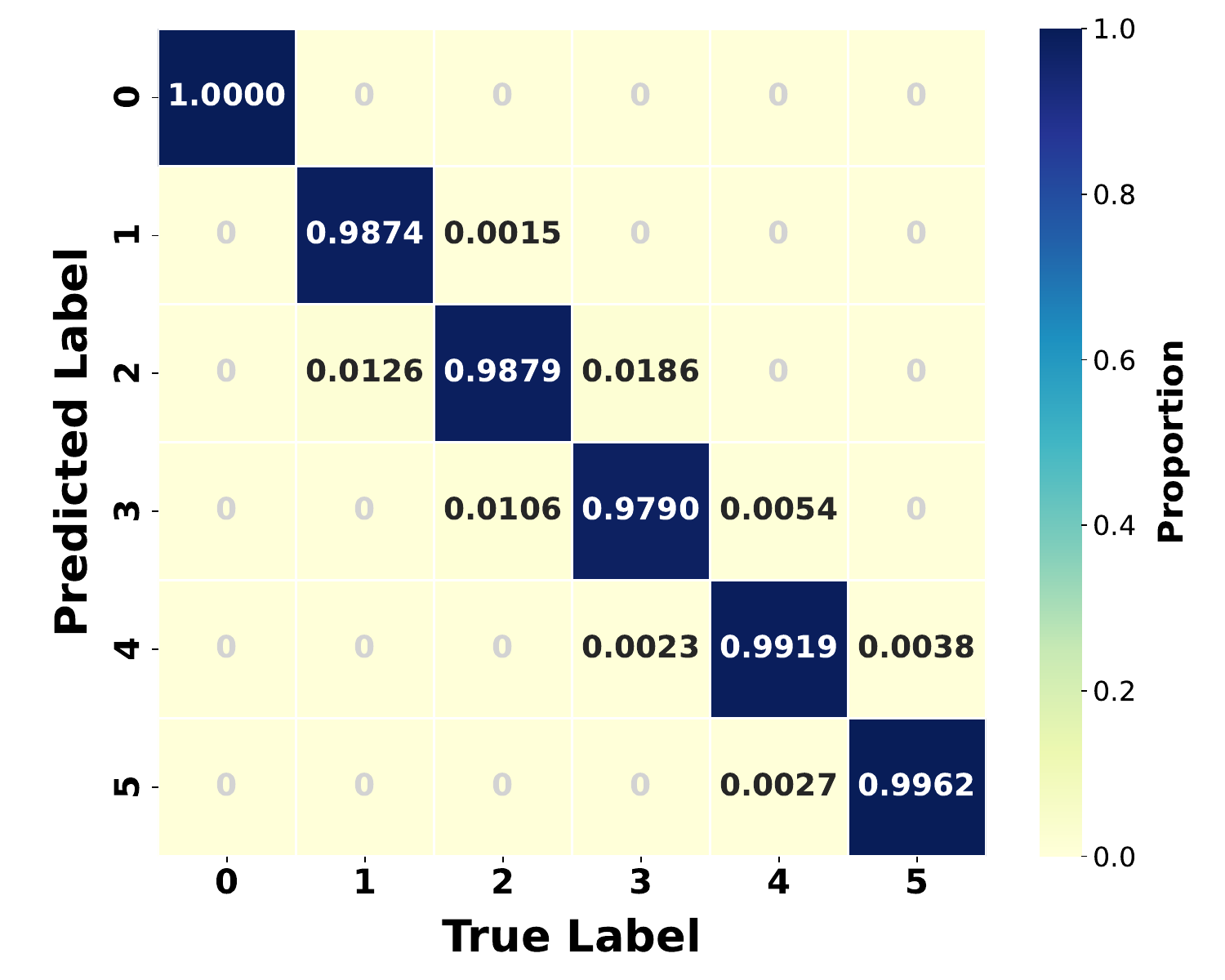}
\caption{Confusion Matrix of the Classification Model}
\label{fig:confusion_matrix}
\end{figure*}

Additionally, we include a confusion matrix in Figure \ref{fig:confusion_matrix} to provide a detailed view of the classification behavior. The confusion matrix reveals that the majority of correctly classified samples are concentrated along the diagonal, indicating strong class discrimination. However, misclassifications are primarily observed along adjacent off-diagonal entries, suggesting that the model occasionally confuses neighboring classes. This behavior is attributed to the natural classification ambiguity arising from the discretization of continuous quantities, where subtle differences between adjacent classes may not always be distinctly captured by the model.

\subsection{Regression Accuracy}
For the regression model focused on predicting bright star counts ($<23.5$ mag), we achieved excellent prediction accuracy across the density range present in our dataset. The model demonstrates strong performance with a mean squared error (MSE) of 0.0178, root mean squared error (RMSE) of 0.1333, and mean absolute error (MAE) of 0.0824 on the normalized logarithmic scale.

Logarithmic metrics are essential given the dynamic range of the data. When tackling regression tasks with a highly skewed and vast dynamic range in target values, such as stellar counts spanning from 0 to over 2000, standard metrics calculated on the linear scale can be disproportionately dominated by large-magnitude errors at the high-end, obscuring the model's performance in critical low-to-medium density regimes. A comprehensive understanding of performance measures dictates that metrics must be tailored to the target variable's distribution to ensure a fair and relevant assessment \citep{terven2025comprehensive}. By calculating MAE and RMSE on the logarithmic scale, we effectively measure the relative prediction error (i.e., the error factor), ensuring that the model is penalized consistently across all density magnitudes.

\begin{figure*}[!ht]
\centering
\includegraphics[width=0.6\linewidth]{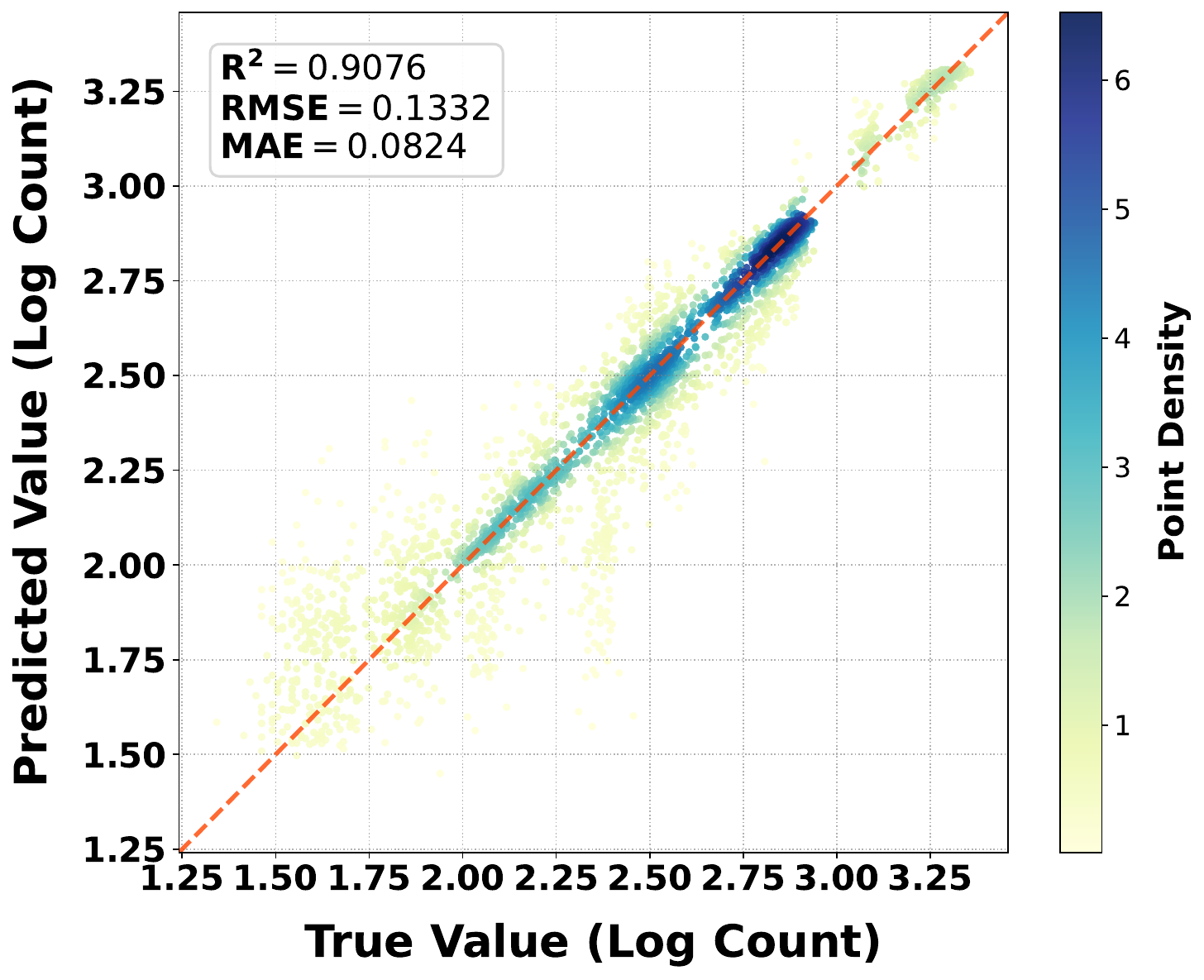}
\caption{Actual versus predicted bright star counts for the regression model. The solid red line represents perfect predictions (y=x), while the dashed lines indicate the $\pm$20\% error margin. The color density reflects the concentration of data points, showing strong agreement across most of the dynamic range with some expected scatter at extreme values.}
\label{fig:actual_vs_predicted}
\end{figure*}

As shown in Figure \ref{fig:actual_vs_predicted}, the scatter plot of actual versus predicted values demonstrates a strong linear correlation along the ideal y=x line, confirming the model's accurate regression capability across the entire density range. The tight clustering of points along the identity line (y=x) indicates consistently reliable estimates, with particularly tight agreement in the medium density regime where most astrometric calibration occurs. Some expected scatter is observed at extreme values, reflecting the inherent challenges in these regions, yet the overall distribution shows excellent predictive performance.

To further interpret these results in practical terms, the RMSE of 0.1333 on the logarithmic scale indicates that the typical prediction error corresponds to a factor of approximately $10^{0.1333}\approx 1.36$ in linear space. This means that for a field containing 100 bright stars, the model's prediction would typically fall between 74 and 136 stars. Similarly, the MAE of 0.0824 translates to an average error factor of $10^{0.0824}\approx 1.21$, indicating that predictions are within about $\pm$21\% of the true values on average.

The high $R^2$ score of 0.9076 indicates that the model explains approximately 90.7\% of the variance in the transformed bright star counts, demonstrating its effectiveness in capturing the underlying density patterns. This strong explanatory power is particularly valuable for astrometric calibration applications, where accurate estimation of available reference stars directly impacts position measurement precision.

The regression model maintains consistent accuracy across the density range covered by our dataset, with particularly strong results in medium density regions where most astrometric calibration occurs. Computational performance remains highly efficient with an average inference time of 0.1ms per image. This exceptional efficiency, combined with the high accuracy, makes the regression model well-suited for integration into the real-time CSST data processing workflow.

To further interpret these results in practical terms, the RMSE of 0.1333 on the logarithmic scale indicates that the typical prediction error corresponds to a factor of approximately $10^{0.1333}\approx 1.36$ in linear space. This means that for a field containing 100 bright stars, the model's prediction would typically fall between 74 and 136 stars. Similarly, the MAE of 0.0824 translates to an average error factor of $10^{0.0824}\approx 1.21$, indicating that predictions are within about $\pm$21\% of the true values on average.

From a scientific perspective, the high coefficient of determination ($R^2=0.9076$) demonstrates that the model captures over 90\% of the variance in bright star counts, directly supporting the pipeline's astrometric stability.By accurately predicting the number of bright stars ($<23.5$ mag), the system can proactively flag fields where astrometric solutions may be ill-constrained (e.g., fields with too few reference sources), allowing for fallback strategies or quality flags in the data products.

The regression model maintains consistent accuracy across the density range covered by our dataset, with particularly strong results in medium density regions where most astrometric calibration occurs. Computational performance remains highly efficient with an average inference time of 0.1ms per image. This exceptional efficiency, combined with the high accuracy, makes the regression model well-suited for integration into the real-time CSST data processing workflow.

\subsection{Misclassification Sample}
\begin{figure*}[!ht]
    \centering
    \subfigure[]{\includegraphics[width=0.3\linewidth]{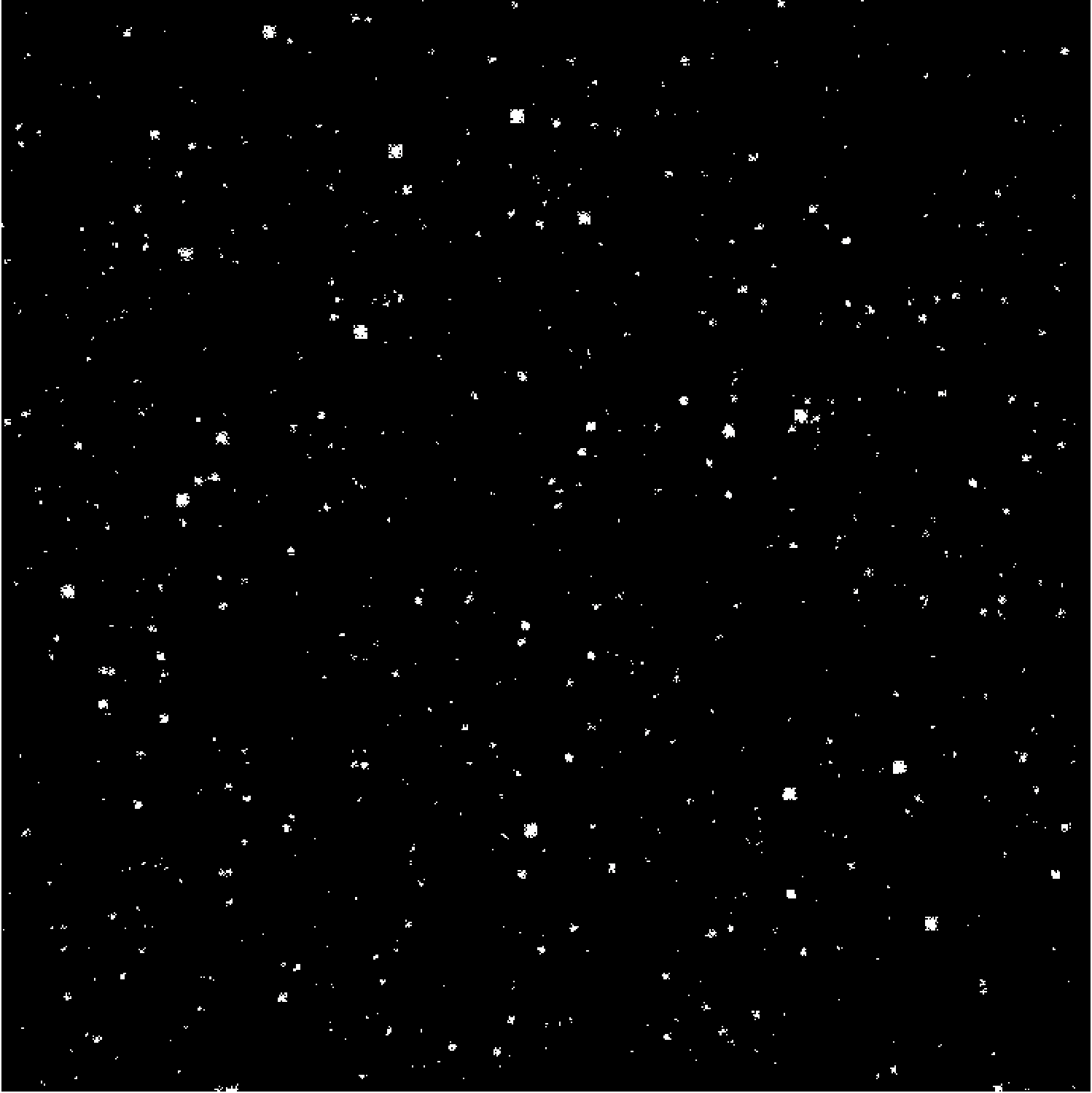}}
    \hfill
    \subfigure[]{\includegraphics[width=0.3\linewidth]{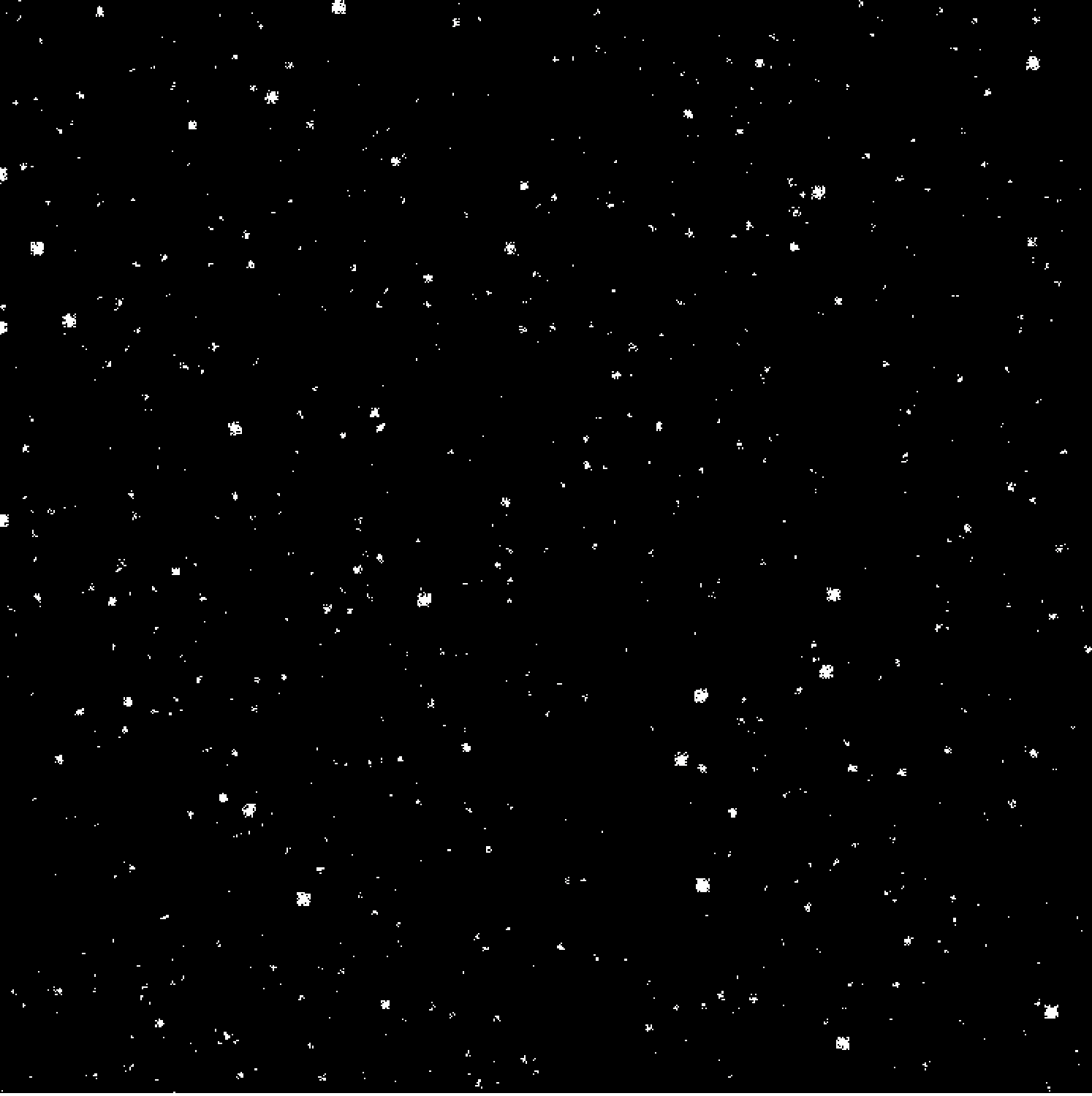}}
    \hfill
    \subfigure[]{\includegraphics[width=0.3\linewidth]{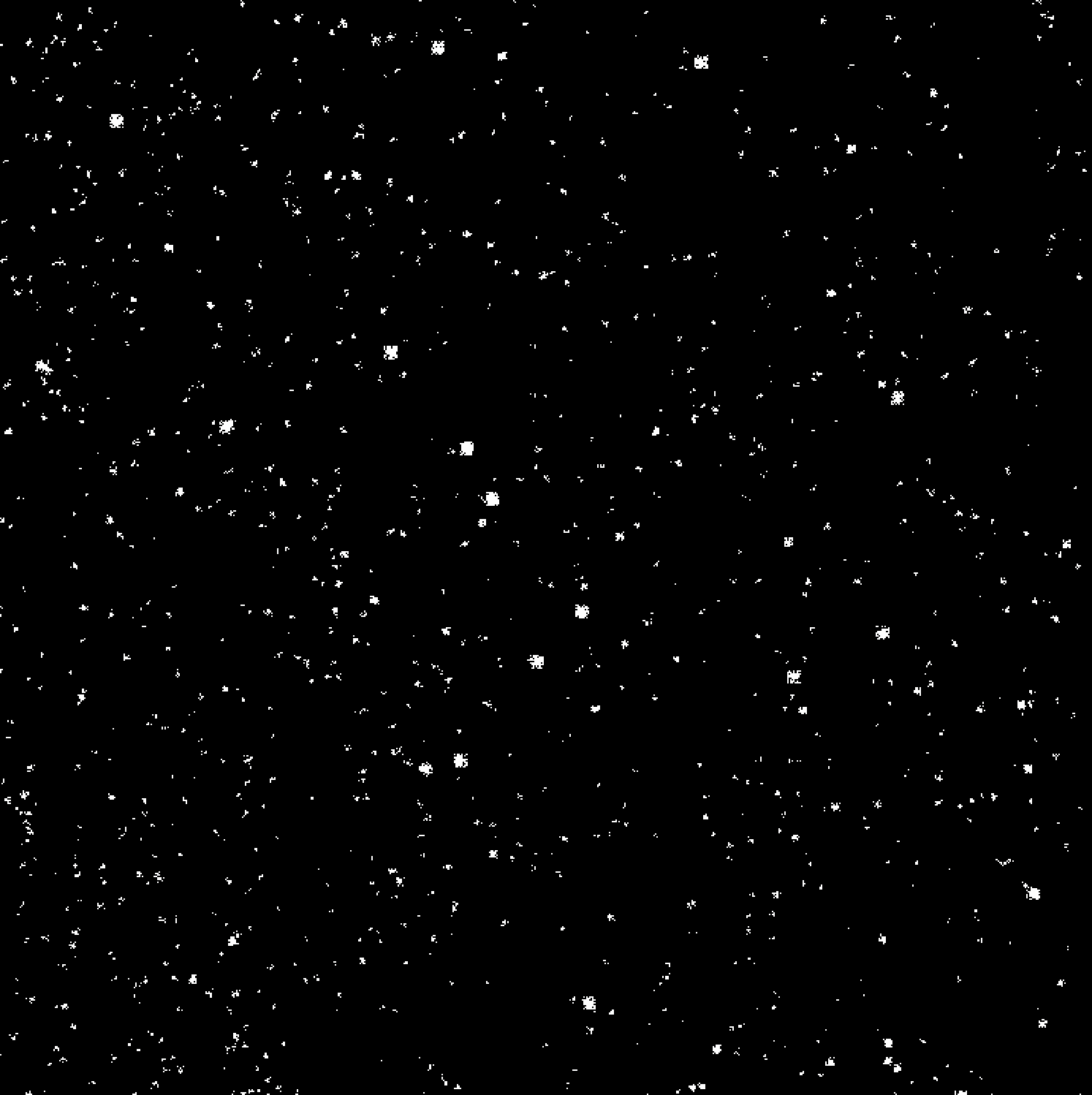}}
    \caption{Density distributions of representative samples: (a) Correctly classified Class 2 sample (966 sources), (b) Class 3 sample misclassified as Class 2 (1005 sources), and (c) Correctly classified Class 3 sample (1343 sources). The similarity between (a) and (b) illustrates the challenge of classifying samples near the 1k-source boundary.}
    \label{fig:error_patterns}
\end{figure*}

To examine the model's classification behavior, we analyze misclassification samples from the test set. Density classes are defined by source counts with thresholds at 0, 100, 1000, 10000, and 100000 sources, spanning Classes 0 to 5, with misclassifications predominantly occurring near these boundaries. Figure \ref{fig:error_patterns} illustrates three representative samples: (a) a correctly classified Class 2 sample (966 sources), (b) a Class 3 sample (1005 sources) misclassified as Class 2 due to its proximity to the 1k-source boundary, and (c) a correctly classified Class 3 sample (1343 sources) far from the boundary. The misclassified sample in Figure \ref{fig:error_patterns}(b) exhibits a density pattern visually similar to Figure \ref{fig:error_patterns}(a), highlighting the morphological ambiguity near the 1k-source threshold that leads to misclassification, while the sample in Figure \ref{fig:error_patterns}(c) is reliably classified due to clearer morphological distinctions.

Crucially, the confusion matrix \ref{fig:confusion_matrix} reveals that misclassifications are almost exclusively confined to adjacent off-diagonal entries. In the context of the CSST pipeline, these constitute 'soft errors' with minimal scientific impact. For instance, the photometric reduction strategy for a high-density Class 2 field is likely compatible with that for a low-density Class 3 field. The model successfully avoids 'catastrophic failures' — such as confusing a sparse extragalactic void (Class 0) with a crowded Galactic center (Class 5)—which would otherwise lead to significant data reduction errors.

\begin{figure}[htbp]
    \centering
    \includegraphics[width=\linewidth]{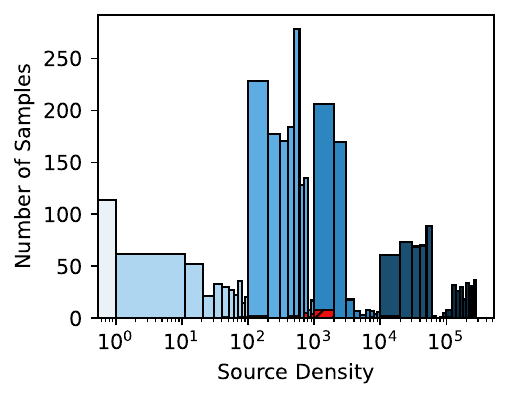}
    \caption{Histogram of source density distribution for all classification test set samples (blue bars) with logarithmically spaced bins, overlaid with the distribution of misclassified samples (red hatched bars). The figure highlights that 50.3\% of samples are concentrated near classification boundaries, with misclassified samples predominantly clustering around the 1k-source threshold for Class 3.}
    \label{fig:Density_distribution}
\end{figure}

This boundary effect is further evidenced in Figure \ref{fig:Density_distribution}, which presents a histogram of the density distribution for all test set samples, with source counts binned logarithmically. The figure shows that 50.3\% of samples across all classes are concentrated near classification boundaries, particularly around the 1k-source threshold. The distribution of misclassified samples, highlighted in red with a hatched pattern, clusters prominently at the Class 3 boundaries (near 1000 sources), confirming that morphological ambiguity in these regions drives the elevated error rate in Class 3.

The discretization of continuous source counts into discrete classes introduces ambiguity at class transitions, particularly for samples like Figure \ref{fig:error_patterns}(b) with a source count (1005) close to the 1000 threshold. The model’s attention mechanisms, while effective for broad density patterns, struggle to resolve subtle morphological differences in these boundary regions, as seen in the near-identical density distributions of Figure \ref{fig:error_patterns}(a) (966 sources) and Figure \ref{fig:error_patterns}(b). To mitigate this, future work could explore adaptive thresholding or continuous density estimation to reduce classification errors in intermediate density regimes like Class 3, improving overall model performance.

For the regression model, while some increased scatter is observed in medium-density regions ($<2.5$ on log scale), this variation does not diminish the model's practical utility. The slightly reduced precision in these transitional densities has minimal impact on pipeline operations, as medium-crowding fields can be adequately processed by standard calibration routines. The model's excellent performance at higher densities fully satisfies the key operational requirement—reliably identifying severely crowded fields that require special processing.
\subsection{Limitation}
While our two-stage framework demonstrates effective density estimation, several limitations should be acknowledged to provide context for our results and guide future improvements.

First, the training data for both models rely on simulated images that cannot fully capture the complexity of real astronomical observations. While our simulations incorporate many instrumental effects, they lack certain astrophysical phenomena such as stray light, cosmic ray impacts, and variable point spread functions that may affect actual CSST observations. Nevertheless, our experimental results confirm that the models have effectively learned meaningful density-related features, providing a solid foundation for future adaptation when actual CSST observational data becomes available.

Second, the sequential nature of our approach introduces potential error propagation between stages. Misclassifications in the initial coarse categorization phase may lead to suboptimal performance in the subsequent regression stage. This architectural limitation could be addressed in future work through confidence-based routing mechanisms or joint optimization techniques that consider both tasks simultaneously.

Finally, both models were trained and evaluated on relatively narrow field-of-view patches, which may miss larger-scale spatial correlations that could be relevant for certain calibration tasks. Expanding the model's receptive field or incorporating multi-scale processing could address this limitation in future implementations.

These limitations, however, do not diminish the practical utility of our current framework, which provides a robust foundation for CSST data processing that can be progressively enhanced as real observational data becomes available.

\subsection{Operational Range}
Defining the valid operational limits of our proposed models is essential for their seamless integration into the CSST data processing system.

The classification model is designed and tested to handle the full dynamic range of CSST observations, from empty fields (Class 0) to extremely crowded galactic centers ($>10^5$ stars per image, Class 5). However, the regression model, which is specifically tailored for astrometric calibration, has a valid range of 0 to about 2000 bright stars ($<23.5$ mag). This upper limit reflects the natural density distribution observed within the 25 deg$^2$ simulation area used for training, rather than an artificial theoretical threshold.

Crucially, this framework is designed to operate on calibrated images, consistent with the CSST pipeline architecture where instrumental artifacts (e.g., cosmic rays, bad pixels) are mitigated by upstream modules prior to source extraction. Therefore, the regression model's effective range and the classification model's dynamic range are aligned with the scientific requirements of the source extraction stage.

Since fields classified as Class 5 ($>10^5$ total sources) are most likely to contain bright star counts exceeding our validated regression range, this classification serves as a trigger to automatically bypass the precise regression stage. Rather than forcing an out-of-distribution prediction, the pipeline can effectively switch to a specialized crowded-field photometry algorithm. This mechanism ensures that the system operates strictly within its verified safety limits, preventing unreliable estimations in ultra-dense regimes.

\section{SUMMARY}
\label{sec:summary}
We have developed a hierarchical two-stage deep learning model to characterize stellar density in multi-color imaging from the Chinese Space Station Survey Telescope (CSST), effectively addressing the challenges posed by its extreme dynamic range (0 to $>10^5$ stars per detector). Our results are listed as follow. 
\begin{itemize}
    \item Among the five different CNN architectures evaluated, which are ResNet-34, Baseline CNN, VGGNet-16, AlexNet, and DenseNet-121, the ResNet-34 demonstrated superior performance in both accuracy (98.83\%) and computational efficiency. Consequently, it was selected as the final model.
    \item The first stage ('classification') employs a ResNet-34 classifier to categorize images into six discrete density regimes with 98.83\% global accuracy, providing a robust and interpretable gate for routing highly crowded fields to specialized deblending algorithms.
    \item The second stage ('regression') utilizes a ResNet-50 regression model to predict the number of bright stars ($<$23.5 mag) essential for astrometric calibration, achieving a mean absolute error of 0.0824 dex.
    \item Error analysis indicates that the first stage misclassifications are primary of 'soft errors' at class boundaries. Meanwhile, the regression model performs reliable with an average error factor of 1.21, and the misclassification mostly occurs in medium-density regions, which does not significantly affect the calibration process. 
\end{itemize}
 

This combined 'classification and regression' model is computationally efficient, fully automated, and readily integrated into any large survey data reduction pipeline. It establishes a foundation for handling heterogeneous density distributions in future wide-field surveys, ensuring higher accuracy and homogeneity of scientific products from Galactic plane to extragalactic fields.

In future work, we plan to fine-tune the model using early commission data from CSST to bridge the gap between simulations and observations, particularly accounting for complex backgrounds and instrumental artifacts. Furthermore, we aim to extend this framework to multi-band joint estimation to improve characterization in heavily reddened regions. This automated tool will eventually serve as a foundational module in the CSST operational pipeline, ensuring high-precision scientific products across the heterogeneous stellar landscape of the upcoming survey.
\section*{ACKNOWLEDGEMENTS}

This work is supported by the National Key R\&D Program of China No.2025YFF0511001. Man I Lam acknowledges the support by National Natural Science Foundation of China (NSFC; grant No. 12373048). Hao Tian acknowledges the support by National Key R\&D Program of 
China No. 2024YFA1611902 and the China Manned Space Program with grant No. CMS-CSST-2025-A11. 

This work is based on the mock data and the softwares created by the CSST Simulation Team, which is supported by the CSST scientific data processing and analysis system of the China Manned Space Project. 

We extend our gratitude to the developers of the ResNet architecture \citep{he2016deep}, which forms the backbone of our classification and regression models, enabling robust feature extraction and efficient training even with highly imbalanced astronomical data. We also thank the PyTorch and torchvision development teams for providing the deep learning frameworks and data augmentation tools that were instrumental in this work.
\bibliography{PASPsample701}{}
\bibliographystyle{aasjournalv7}

\end{document}